# Electrically-driven amplification of terahertz acoustic waves in graphene

Aaron H. Barajas-Aguilar[1], Jasen Zion[2], Ian Sequeira[1], Andrew Z. Barabas[1], Takashi Taniguchi[3], Kenji Watanabe[3], Eric Barrett[1], Thomas Scaffidi[1], Javier D. Sanchez-Yamagishi[1].

[1]Department of Physics and Astronomy, University of California, Irvine, Irvine, CA, USA, [2]T. J. Watson Laboratory of Applied Physics, California Institute of Technology, Pasadena, California, USA, [3]Research Center for Functional Materials, National Institute for Materials Science, 1-1 Namiki, Tsukuba, Japan.

**Abstract**

In graphene devices, the electronic drift velocity can easily exceed the speed of sound in the material at moderate current biases. Under this condition, the electronic system can efficiently amplify acoustic phonons, leading to the exponential growth of sound waves in the direction of the carrier flow. Here, we demonstrate that such phonon amplification can significantly modify the electrical properties of graphene devices. We observe a super-linear growth of the resistivity in the direction of the carrier flow when the drift velocity exceeds the speed of sound, causing up to a 7 times increase over 8 micrometers. The resistance growth is observable for carrier densities away from the Dirac point and is enhanced at cryogenic temperatures. These observations are explained by a theoretical model for the electrical-amplification of acoustic phonons, which reach frequencies up to 2.2 terahertz with the nanoscale wavelength set by gate-tunable ~$k_F$ transitions across the Fermi surface. These findings offer a route to high-frequency on-chip sound generation and detection, which can be used to modulate and probe electronic physics in van der Waals heterostructures in the terahertz frequency range.

**Main**

Sound waves are important as high-frequency signal carriers and as a means to distort crystal lattices in space and time. Due to the slow speed of sound compared to light, ultrashort sound wavelengths in the nanometer scale are attainable in the terahertz (THz) domain, the highest sound frequencies in solids. The control and generation of THz sound waves offers a route to nanoscale ultrasonic imaging, the generation of THz electromagnetic radiation[1–3], and the dynamic modulation of electronic behaviors[4–7]. However, an electrical on-chip source of THz acoustic waves remains elusive. Coherent THz sound waves have only been achieved via ultrafast optical pumping[8,9], while conventional piezoelectric transducers produce maximum frequencies of ~ 10 GHz[10].

Accelerated electrons can generate and amplify traveling waves. Famous examples include the traveling wave amplifier[11] and Cherenkov radiation. The common gain condition is that the electron velocity exceeds the wave phase velocity. In solids, an analogous effect occurs when the carrier drift velocity ($v_D$) exceeds the speed of sound ($v_s$), resulting in acoustic wave amplification[12,13]. Unlike the free electron case, the amplification has a characteristic wavelength given by transitions across the Fermi surface (Fig. 1a). Such acoustic amplification has been studied in bulk semiconductors[14,15], and has recently been used to make nonreciprocal acoustic amplifiers operating in the gigahertz frequency range[16].

Two-dimensional van der Waals (vdW) materials present many advantages for acoustoelectric devices because phonons are naturally confined to atomic layers, leading to long lifetimes[17] and more efficient coupling to electrons[18]. Acoustic waves offer a way to dynamically modulate lattice strain in both space and time, which can couple to diverse vdW heterostructure phenomena[19], as well as to quantum defects[20]. Despite this interest, acoustic studies in vdW materials have been limited by the challenge of generating propagating sound waves at high frequencies, as the confined nature of the two-dimensional (2D) phonons makes it difficult to excite with external transducers.

Graphene is an attractive host for the electrical amplification of acoustic phonons. Its high Fermi velocity and electron mobility ($\mu$) means that large drift velocities ($v_D=\mu E$) can be achieved at relatively small electric biases. Under a current bias, the drifting Fermi distribution results in an effective population inversion with an energy difference $\Delta E=\hbar v_D 2k_F$ between forward and backward propagating carriers (at zero temperature), where $k_F$ is the Fermi wavevector. When $v_D>v_S$ ($v_{S-TA}$=13 km/s and $v_{S-LA}$=21 km/s for TA and LA phonons, respectively[21]), electrons can relax energy and momentum by emitting and amplifying waves via inelastic backscattering (Fig. 1a). Importantly, over a large range of parameters, the only excitations that graphene can generate are acoustic waves[22,23], which are in the THz frequency range. Evidence for acoustic phonon amplification has come from noise measurements in graphene devices[24]. However, the direct effects of THz acoustic waves on the electronic properties of materials are still unexplored.

Here, we study the transport behavior of clean graphene devices as a function of voltage bias at cryogenic temperatures. In contrast to previous bias studies[25–27], we measure the position-dependent resistance, and focus on moderate voltage biases away from the Dirac point so that interband transitions and optical phonon generation is suppressed. Our primary findings are that graphene resistivity grows strongly in the direction of carrier flow when the drift velocity exceeds the speed of sound. Our results are well explained by electrically-induced amplification of terahertz acoustic waves and their associated strong scattering of graphene electrons.

To study the spatial dependence of the graphene resistance, we fabricated long graphene devices encapsulated in hexagonal boron nitride (hBN) with equally spaced voltage probes along the channel length (inset Fig. 1c). A DC source-drain voltage bias is applied to the device, and by measuring the voltage difference between adjacent electrodes we probe the spatial distribution of the voltage drop across the channel as a function of the current flowing between the source and drain electrodes. The resulting V-I curves are shown in Figure 1c (top panel), measured at an electron carrier density n=1.4×10$^{12}$ cm$^{-2}$ and cryostat temperature of 1.5 K. At low currents, all curves show linear Ohmic behavior corresponding to an average resistivity of 19.7 Ω/square with less than 6% deviations across the sample. The average carrier mobility is 2.3x10$^5$ cm$^2$/V*s. At higher current magnitudes, all the curves deviate strongly from Ohmic behavior with a differential resistance (dV/dI) that grows superlinearly with the magnitude of the source-drain current (Fig. 1c, bottom panel).

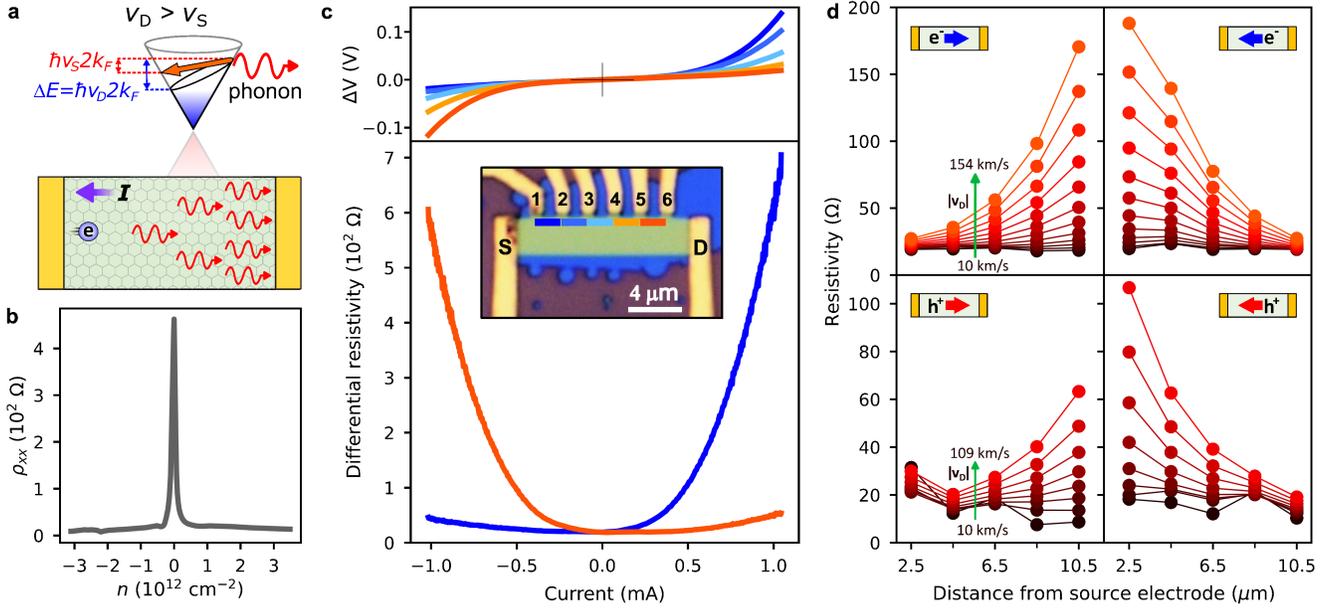

**Figure 1: Acoustic-phonon amplification and observation of resistance growth in direction of carrier flow. a)** Schematic of the acoustic phonon amplification process in graphene. The energy difference between right and left moving carriers is $\hbar v_D 2k_F$, when $v_D>v_S$ the carriers can backscatter and emit a phonon of energy $\hbar v_S 2k_F$. **b)** Resistivity versus electron carrier density for device A, the absence of satellite peaks indicates that the graphene and hBN layers of the device are unaligned. **c)** top: Voltage difference (ΔV) between pairs of consecutive contacts vs. source-drain current for device A. The voltage differences $\Delta V_{1-2}$ (blue) and $\Delta V_{5-6}$ (orange) exhibit the largest non-Ohmic behavior (n=1.4×10$^{12}$ cm$^{-2}$), bottom: differential resistivity vs. source-drain current for the outermost pairs of contacts ($\Delta V_{1-2}$ and $\Delta V_{5-6}$) showing asymmetric non-Ohmic behavior. Inset: optical image of device A with a 13 µm length, 3 µm width and center to center distance between voltage tabs of 2 µm. The colored bars label the pairs of contacts used to measure the voltage differences plotted in the top and bottom panels, **d)** Position dependence of the resistivity at different drift velocities for n=+1.4×10$^{12}$ cm$^{-2}$ (top panels) and n=-1.4×10$^{12}$ cm$^{-2}$ (bottom panels). The device cartoons show the carrier flow direction and carrier type for each case. The maximum drift velocity in the top panels is 154 km/s (j=0.34 mA/µm), and 109 km/s in the bottom panels. Note that the maximum $v_D$ achieved for holes is lower than for electrons due to larger source-drain contact resistance. All measurements are performed at T = 1.5 K.

Strikingly, the nonlinear resistance is highly asymmetric with the current direction and strongly position dependent. The largest asymmetry and nonlinearity are found for the measurements closest to the source and drain electrodes (blue and orange lines Fig. 1c). For example, for contacts 1-2, the differential resistivity at -1 mA is 44 Ω, but at +1 mA it rises to 612 Ω, a factor of 14x difference for opposite current directions. Measurements on the opposite side of the device (5-6) produce similar nonlinear curves, but with opposite dependence on the current direction.

The position dependence of the graphene resistivity shows superlinear growth in the direction of carrier flow (Fig. 1d): for electron-doping, the resistivity growth is opposite to the current; for hole-doping, growth is in the direction of the current. The growth is substantial at moderate current densities j=0.34 mA/µm, leading to a 7x increase in resistivity and 12x increase in differential resistivity over 8 μm (Fig. S1). As a result of the highly nonuniform resistivity at this bias, 81% of the graphene resistance occurs in the last 46% of the channel length (Fig.S2).

The nonlinear resistance growth occurs for a wide range of carrier densities. This is seen in Figure 2a, where, for contacts 1-2 we observe strongly asymmetric dV/dI versus current curves for carrier densities ranging from $0.6 \times 10^{12}$ to $3.5 \times 10^{12}$ cm$^{-2}$. For all densities, the resistance is larger in the direction where the carriers travel longer distances in the device (downstream, Fig. 2a).

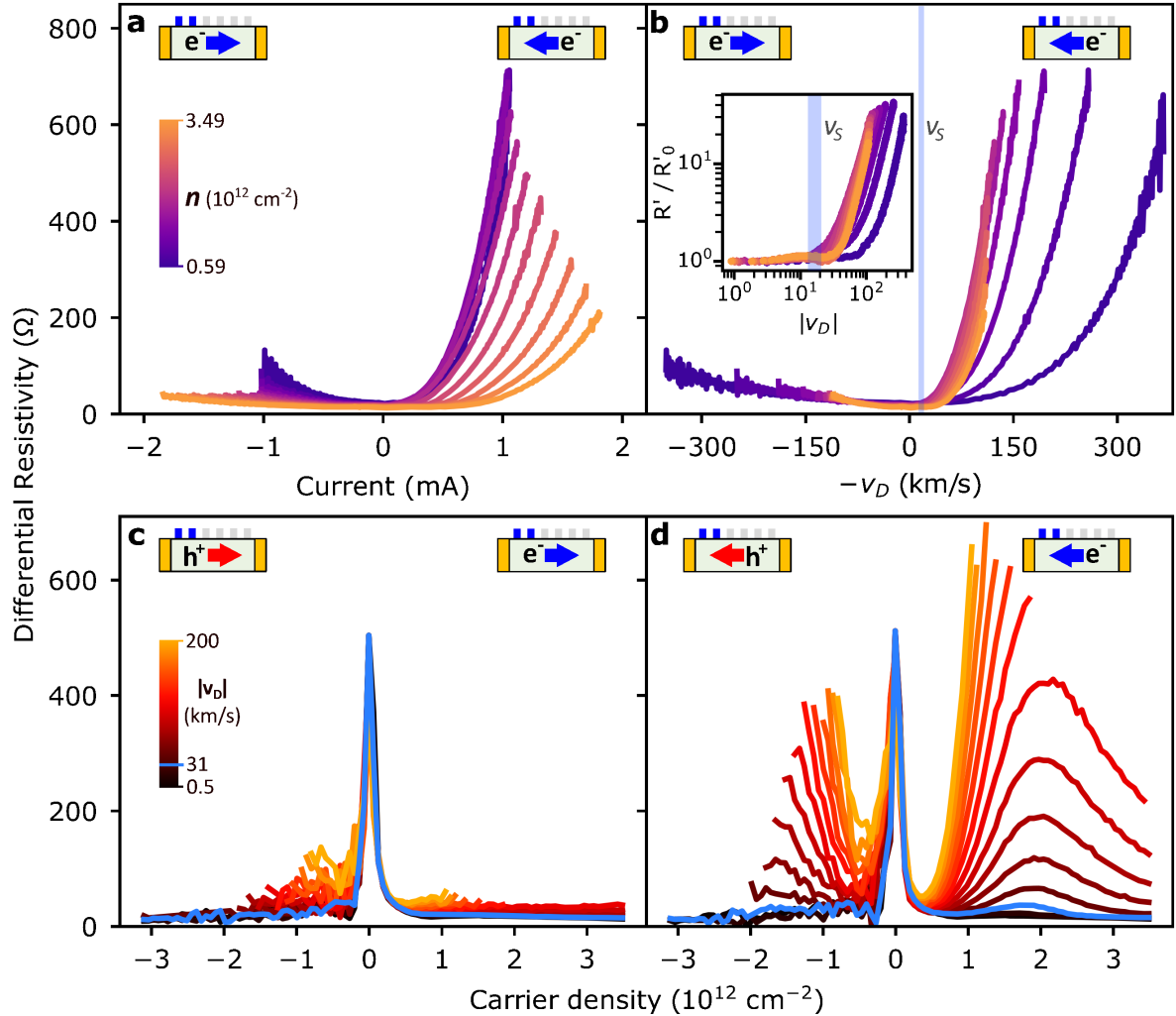

**Figure 2. The resistance growth is sensitive to carrier-density and only occurs when $v_D > v_S$. Top:** Differential resistivity as a function of current **(a)** and drift velocity **(b)** at different carrier densities. The shadowed regions indicate drift velocities between 13 and 21 km/s, corresponding to the speed of sound for TA and LA phonons respectively. Inset: Logarithmic plot of differential resistivity normalized to the value at $v_D=0$. Bottom: Carrier density dependence of the resistivity at positive and negative drift velocities (**c** and **d**, respectively). The device cartoons indicate the carrier type and flow direction, as well as the contacts being measured in each case (contacts 1-2 for all the panels). The highlighted blue traces correspond to a drift velocity of 31 km/s, which is the lowest drift velocity above $v_{S-LA}=21$ km/s shown in this plot. All measurements are performed at T = 1.5 K.

When plotting dV/dI versus drift velocity ($v_D$=j/ne) (Fig. 2b), the curves with carrier densities above $1.1×10^{12}$ cm$^{-2}$ collapse together, suggesting that the physics of this phenomena is dictated by a drifting electronic carrier distribution. On a logarithmic scale plot (Fig. 2b inset), the normalized differential resistivity versus $v_D$ shows a sharp transition from constant to a growing non-Ohmic behavior. The threshold drift velocities, which we define as a 1.5x increase in differential resistivity, vary with carrier density from 21.8 to 83.3 km/s. Notably, non-Ohmic behavior is only observed above the lowest graphene sound velocity ($v_{S-TA}$=13 km/s). The sharp transition between the Ohmic and non-Ohmic regimes can be observed for any pair of contacts in the device (Fig. S3).

When the voltage drop is measured closest to the carrier injection point (upstream, Fig. 2c), the resistance versus carrier density curves follow the typical graphene Dirac peak response with a weak dependence on drift velocity, indicating mostly Ohmic behavior. When the carrier flow is reversed, the carriers travel 10.5 µm before reaching the measuring contacts (downstream, Fig. 2d). Here, for drift velocities lower than $v_s$, the line traces show a typical graphene response - decreasing in resistance with increasing carrier density magnitude (Fig. 1b). But, when $v_D$ is larger than the speed of sound (light blue line), the differential resistivity instead grows rapidly with carrier density for n larger than $0.35×10^{12}$ cm$^{-2}$, surpassing the value of the resistance at the Dirac peak. This effect can be seen symmetrically for both electron and hole carriers. As the carrier density increases, we observe a peak in the dV/dI at n ~$2×10^{12}$ cm$^{-2}$, and an eventual downturn at higher electron doping. Similar behavior is observed for all the other pairs of contacts along the device (Fig. S4).

The resistance growth is most prominent at cryogenic temperatures. Figure 3a displays the differential resistivity versus drift velocity measured for contacts 1-2 from T=1.5 to 280 K. At T=1.5 K, a highly asymmetric curve is obtained with a large nonlinear resistivity growth when measuring far from the carrier injection point (downstream, negative $v_D$). As the temperature increases, the nonlinearity and asymmetry are steadily reduced, but are still evident even at T=280 K. Correspondingly, the resistance dependence on temperature shows an opposite trend for different current bias directions (Fig. 3b). In an upstream configuration (top panel Fig. 3b), a steady increase of the resistivity with temperature is observed as is typical for graphene at low biases, consistent with increased scattering from thermally-occupied acoustic phonons[28,29]. For the downstream measurement (bottom panel Fig.3b), a similar behavior is observed for small drift velocities. But, as $v_D$ increases beyond 37 km/s, the opposite dependence is observed, with resistivity decreasing as the temperature rises above 40 K. At 1.5 K, the effect of drift velocity on resistivity is 3.5 times greater than the effect of heating from 1.5 K to 280 K.

To summarize, we observe the graphene resistivity to grow superlinearly in the direction of carrier flow when the drift velocity exceeds the speed of sound. The resistance growth is suppressed at low carrier densities, and it is strongest at cryogenic temperatures. These observations were made in two graphene devices where the graphene is misaligned with its encapsulating hBN layers (Figs. S6, S7 and S8). The directional resistance growth is not observed in devices where the graphene and hBN form a moiré superlattice (Fig. S9), or where there is too much disorder (Fig. S10).

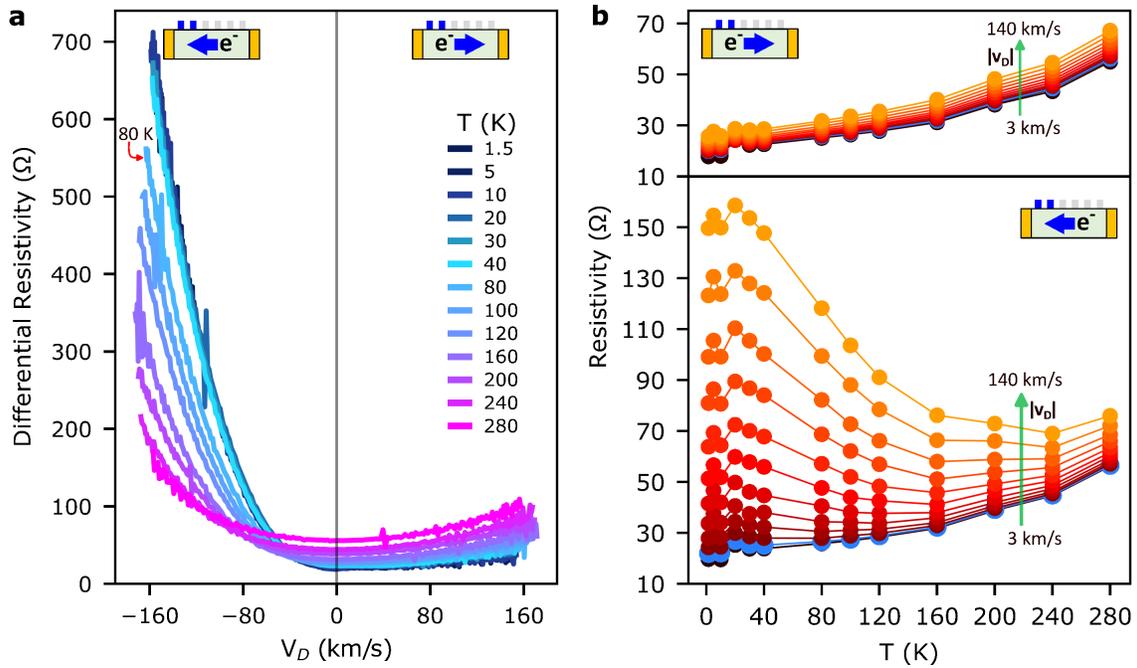

**Figure 3. Resistance due to amplified acoustic phonons is largest at low temperatures.** a) Differential resistivity vs. drift velocity measured for contacts 1-2 at different temperatures from 1.5 to 280 K at n=1.4×10$^{12}$ cm$^{-2}$. (b) Resistivity for contacts 1-2 as a function of temperature for constant drift velocities. The green arrows indicate the direction of growth of the $v_D$ magnitude. The blue traces correspond to $v_D$=26 km/s, the lowest $v_D$>$v_{S-LA}$ shown in this plot. The temperature dependence inverts at high drift velocities. The excess resistivity induced by high $v_D$ at 1.5 K is 130 Ω. The excess resistivity induced by heating from 1.5 K to 280K for $v_D$=3 km/s is 37 Ω.

## Discussion.

In low-disorder conductors, acoustic phonons are the simplest low energy excitation that can relax the momentum of electrons accelerated by electric fields and create resistance. At low temperatures, this form of dissipation is only unlocked when the drift velocity exceeds the speed of sound, due to energy-momentum conservation. This can be understood from the form of the tilted Fermi distribution in a drifting rest frame , where the effective population inversion between forward and backward moving carriers enables electrons to backscatter by emitting acoustic phonons in the direction of the carrier flow, with characteristic wavevectors $k_{phonon}$~$2k_F$ and energy $\hbar v_s k_{phonon}$ (Fig. 1a).

In this work, the range of drift velocities where the resistivity growth is observed points to an acoustic-phonon mechanism. The threshold behavior for the drift velocity is particularly sharp, with little change to the local resistivity (< 15%) and highly symmetric V-I curves until the sound velocity is exceeded (Fig. 2b). When $v_D$= $v_{S-LA}$, the tilt of the Fermi distribution is $\hbar v_D 2k_F$ = 5.8 meV at n=1.4×10$^{12}$ cm$^{-2}$. At these low energy scales, acoustic phonons are the only excitation available for inelastic energy transfer. At the highest drift velocities that we probe at such carrier density (160 km/s), the energy tilt is 43 meV and is not sufficient to directly excite the lowest energy optical phonons of the device (102 meV for hexagonal boron nitride[30]).

Typically, electron-phonon scattering is considered as solely a source of dissipation, leading to local heating of the electron and phonon distributions. But such Joule heating effects would not produce the strongly asymmetric resistance profile that we observe (see Methods section on heating effects). Instead, if the acoustic phonons are long-lived, they will propagate downstream and stimulate the emission of additional phonons, producing an exponential growth in the direction of carrier flow. The only condition for growth is that the net rate of phonon emission exceeds the phonon decay rate, the latter being small at cryogenic temperatures[31]. The phonon population growth will be mirrored in the resistivity, as each phonon emission occurs with an electron backscattering. Thus, acoustic-phonon amplification results in an exponentially growing resistance in the direction of carrier flow when $v_D > v_S$.

We calculate the phonon amplification rates and effects on the graphene resistance using a model of the driven electron-phonon dynamics across the channel length (See Methods and SI section 5). We find a cone of phonon modes in the direction of the drift velocity that will be amplified resulting in exponential growth in the direction of the carrier flow as , where $\Gamma_k$ is the amplification rate for mode k, and x is the position. $\Gamma_k$ reaches levels of 1 to 40 GHz for $v_D$ values from 2 to 10*$v_{S-LA}$, with a broad maximum near $k=k_F$ along the x direction (Fig.S11). Extending beyond previous works, we calculate the spatial growth rate of the resistivity, and find it to be well approximated by the peak value of $\Gamma_k/v_S$ with values of 0.3 to 2.5 µm$^{-1}$ (Fig. S12). Such micron-scale growth lengths are comparable with our experimental observations. For the data measured at $v_D$=7.33*$v_{S-LA}$ (154 km/s), n=1.4×10$^{12}$ cm$^{-2}$, the observed trend is well fit by an exponential with a characteristic growth rate ~0.32 µm$^{-1}$ (Fig. S2). This value is 5x less than the theoretical calculation, which is not surprising given that the model neglects phonon loss mechanisms such as anharmonic decay and edge scattering.

The nonmonotonic dependence of the resistivity growth on carrier density can also be understood within the phonon amplification model. Near the Dirac point, we expect phonon amplification to be suppressed due to competing pathways for energy relaxation via interband excitations, such as electron-hole generation, when the Fermi tilt is comparable to the Fermi energy[27]. Moreover, the electron-phonon coupling and amplification rates are reduced with smaller $k_{phonon} \sim k_F$ (Fig. S11). Conversely, as the carrier density increases, the larger $k_{phonon}$ will have stronger electron-phonon couplings and subsequent higher rates of amplification and resistive scattering with electrons[24,28]. The larger Fermi surface also amplifies a larger range of modes, further increasing the resistivity growth. An opposing effect is the increase of the phonon decay rates at larger phonon wavevectors[32], either due to anharmonic decay or short-range disorder, which will lower the amplification rates at higher carrier densities. Our data is in agreement with these aspects of the phonon amplification model, where we observe a sharp increase in the resistivity growth as we dope away from the Dirac point, a maximum at n~2×10$^{12}$ cm$^{-2}$, and a downturn for higher n values (Fig. 2d).

The requirement that the phonon emission rate exceeds the decay rate for net amplification makes the process also sensitive to the overall lattice temperature. The phonon decay rate will increase with thermal mode occupation due to anharmonic processes, which will suppress the amplification process. Indeed, the resistance growth is strongest at T=1.5 K and is only suppressed when the temperature exceeds the energy scale of the amplified modes ($\hbar v_S 2k_F$=40 K for n=1.5×10$^{12}$ cm$^{-2}$) (Fig. 3b). The unique end result is a metallic conductor with a larger resistance at cryogenic temperatures than at room temperature when biased.

Other processes that scatter acoustic phonons will also suppress phonon amplification. This explains why we do not observe the resistance growth in disordered devices (i.e., with low electronic mobility, Fig. S10), or with a moiré superlattice potential (Fig. S9). In the case of aligned graphene-hBN devices, the moiré superlattice can be as large as 14 nm, which will scatter the small wavelength acoustic phonons[33].

In summary, we demonstrate an unprecedented directional growth of the graphene resistance induced by electrically-amplified acoustic phonons. Our results have important implications for graphene applications, especially at high current densities or over long distances where phonon amplification is likely to be a limiting factor. At the same time, these observations show the unique potential of high-frequency acoustic waves to remotely modulate the electrical properties of a material. The strong modulation of the graphene resistance is only possible due to the large wavevectors of the generated phonons, which can backscatter electrons across the Fermi surface. Low wavevector phonons with $k_{phonon}$<< $2k_F$, as would be predominantly generated by a thermal pulse, can only induce small-angle scattering and hence weakly affect the resistivity. For similar reasons, traditional acoustic-electronic studies, which use surface acoustic waves with wavelengths >> 100 nm, can only act as slow and long length-scale perturbations for carrier densities above 1×10$^{11}$cm$^{-2}$ ($\lambda_F$ < ~100 nm). In our devices, we estimate that we are predominantly amplifying acoustic phonons with wavelengths of 40 nm down to 9 nm. Interestingly, such length scales are comparable to the moiré superlattices found in twisted van der Waals heterostructures[34,35], motivating future work studying the dynamic spatiotemporal strain effects of acoustic waves.

The characteristic frequencies of the amplified acoustic phonons are in the terahertz range (0.3-2.2 THz) and are tunable with the graphene Fermi energy $E_{phonon}$ ~ $(v_S/v_F)E_F$. Currently, there are no alternative demonstrations of electrical generation of terahertz acoustic phonons. Transducing the mechanical motion of the acoustic wave to an electric field would offer a route towards a terahertz electromagnetic source. Acoustic amplification should be observable in other high-mobility vdW materials such as transition metal dichalcogenides [36], where intrinsic piezoelectricity can convert the sound waves into electric waves. Lastly, Cerenkov phonon amplification in graphene offers a unique route to the electrical generation of other high-frequency and large wavevector excitations in 2D heterostructures, such as acoustic plasmons[37] and magnons[38], which are otherwise challenging to source and probe.

## Methods

### Fabrication
All graphene and hBN layers were exfoliated from bulk crystals. Stacks were fabricated by the dry transfer method[39] using stamps made of a polycarbonate (PC) film on top of a polydimethylsiloxane (PDMS) square on a glass slide. All the lithographic processes were made by electron beam lithography (EBL) using a layer of poly(methyl methacrylate) (PMMA) resist. To write the patterns for the one-dimensional (1D) edge contacts, PMMA 950 A5 was spun for 2 minutes at 2000 rpm producing a ~500 nm thick layer. The EBL patterns were written at 1.6 nA with 30 kV excitation and then developed for 2 to 3 minutes in a cold mixture of 3:1 isopropyl alcohol (IPA)/water. After writing and developing the patterns, reactive ion etching (RIE) was used to expose the graphene with the following parameters: a flow of 10 standard cubic centimeters per minute (SCCM) of SF6, 2 SCCM of O2 and 30W of radio frequency power, at 100 mTorr for 30 s[40]. Then, 3 nm of Cr and 100 nm of Au were deposited in an electron beam evaporator system at 1 Å/s. Liftoff was performed by soaking the sample in acetone for 1 to 2 hours and agitating with a pipette. To define the devices geometry a mask was written with EBL and finally a two-step RIE process was made, first an SF6 etching using the same parameters described for the 1D edge contacts, and then an $O_2$ etching with a flow of 20 SCCM of $O_2$, 30 W of radiofrequency power at 70 mTorr for 15 s.

**Device Measurements details**

The devices were measured in a variable temperature cryostat. For the transport measurements, a source-drain DC voltage bias was applied to the devices while measuring the source current using a Keithley 2400 SMU. The voltage drop between consecutive pairs of contacts was measured using digital multimeters. A gate voltage was applied to the silicon backgate to control the carrier density. To calibrate the gate capacitance value for device A in the main text, which determines the calculated values of the drift velocity, the Landau fan features up to B = 3T were measured and fitted. Using the measured capacitance and thicknesses of the dielectric layers, we extract a value for the out-of-plane dielectric constant of hBN $\varepsilon_\parallel$=3.44, which agrees with the reported value[41]. This dielectric constant value for hBN was used in the data analysis for the rest of the devices.

**Theory Calculations**

To calculate the resistance in a long graphene channel under a current bias, we assume a drifting Fermi-Dirac distribution with drift velocity $v_D$ for the electrons in order to calculate the phonon amplification rate due to electron-phonon coupling via the deformation potential. We then use to calculate the position-dependent out-of-equilibrium phonon distribution in the sample. Finally, we find the position-dependent electric field needed to sustain the assumed $v_D$ based on the electronic Boltzmann equation in which the phonon distribution enters through the electron-phonon scattering integral.

Our primary assumption is that of a drifting Fermi-Dirac distribution for the electrons, with a nominal temperature which is $v_D$-independent. In the actual system, we would expect that Joule heating would lead to increased electronic temperatures at high $v_D$. We also neglect phonon loss mechanisms such as disorder scattering or anharmonic decay, which will limit the phonon amplification process at large temperatures and large phonon population levels. As such, the model is best in capturing the qualitative aspects of phonon amplification, especially at lower drift velocities. For a detailed explanation of the theoretical calculations see Section 5 in the supplementary material.

**Relation to previous noise studies of acoustic-phonon amplification in graphene**

A previous work by Andersen et. al.[24] provided evidence for acoustic-phonon amplification in graphene via noise measurements and by extracting phonon transit timescales from AC transport measurements. Our study is differentiated by showing how acoustic-phonon amplification can directly modify the graphene DC resistance, leading to dramatic and easily measurable effects that are spatially varying.

**Other mechanisms for spatially-dependent resistance at high bias**

Large source-drain voltage biases can lead to a spatially-dependent carrier density across the channel[26]. This can lead to a spatially varying device resistance which is significant if the voltage bias is comparable to the gate voltage. In our measurements, the maximum applied biases are 0.6 V, leading to a Δn ~ 4×10$^{10}$ cm$^{-2}$ change in the carrier density across the device channel. This is small (1%) compared to the much larger gate voltages and carrier densities which are the focus of the presented data ($V_{Gate}$ = ±50 V and n~±3.3×10$^{12}$ cm$^{-2}$), and thus cannot explain the observed resistance growth.

Joule heating due to resistive dissipation can lead to a spatially-dependent temperature profile with corresponding spatially-dependent resistances and thermovoltages. In a diffusive system, Joule heating depends only on the magnitude of the current, leading to a temperature profile that is symmetric with the current. The long devices that we measure are in the diffusive regime, as the electron mean free path is substantially smaller than the channel length due to edge scattering (sample width = 3 μm, length = 13 μm), and will get smaller as the electron-acoustic phonon scattering increases when $v_D$>$v_S$. Thus, in our devices Joule heating is expected to produce a symmetric temperature profile and cannot explain the strongly asymmetric resistance profile we observe.

## Acknowledgements.

The authors acknowledge the use of facilities and instrumentation at the Integrated Nanosystems Research Facility (INRF), in the Samueli School of Engineering at the University of California Irvine, and at the UC Irvine Materials Research Institute (IMRI), which is supported in part by the NSF MRSEC through the UC Irvine Center for Complex and Active Materials. The authors also acknowledge the use of the UCI Laser Spectroscopy Lab. The authors thank L. Jauregui, V. Fatemi, and E. Pop for productive discussions, as well as the technical assistance of Q. Lin, R. Chang, M. Kebali, J. Hes, and D. Fishman.

**Funding:** This work was partially supported by National Science Foundation Career Award DMR-2046849. AHBA acknowledges the University of California Institute for Mexico and the United States (UC MEXUS) for partial financial support. IS acknowledges fellowship support from the UCI Eddleman Quantum Institute.

# Supplementary Material.

Contents

# 1. Additional data on device A.

## 1.1 Distance dependence of differential resistivity.

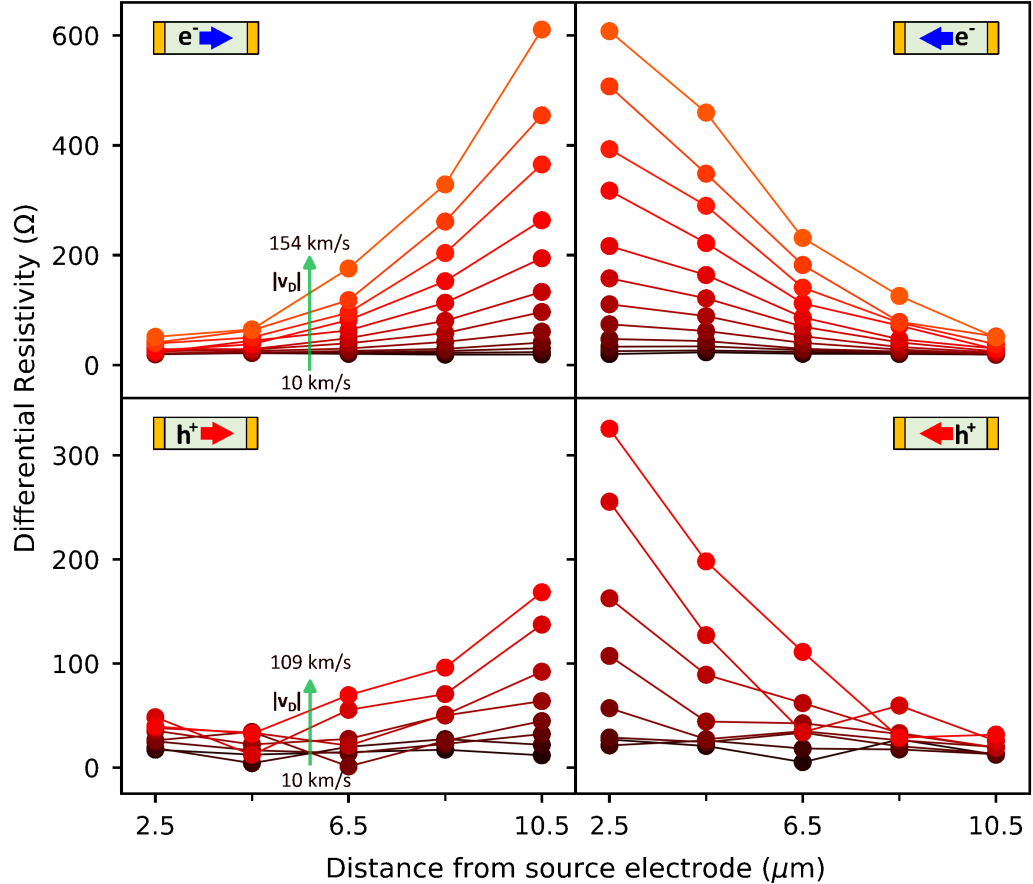

**Figure S1. Distance dependence of differential resistivity** for device A at carrier densities of $1.4 \times 10^{12}$ cm$^{-2}$ (top panel) and n=$-1.4 \times 10^{12}$ cm$^{-2}$ (bottom panel). In the top panels the maximum current density (j=0.34 mA/µm) corresponds to a $v_D$=154 km/s and a 12x increase of the differential resistivity in 8 µm can be appreciated. For the bottom panels the maximum $v_D$ value is 109 km/s.

## 1.2 Exponential resistivity growth with distance and total resistance calculation.

The resistivity vs. distance curve for $v_D$=154 km/s for right-moving electrons (top-left panel of Fig.1d) was fitted by an exponential function of the type $r = m + ae^{(b*x)}$ (Figure S2). The parameters a, b and m were fitted using a nonlinear least squares method. From this fit, we calculate the total resistance of the graphene channel to be 481 Ω at $v_D$ = 154 km/s, where 81% of that resistance comes from the last 6 µm of the 13 µm long device.

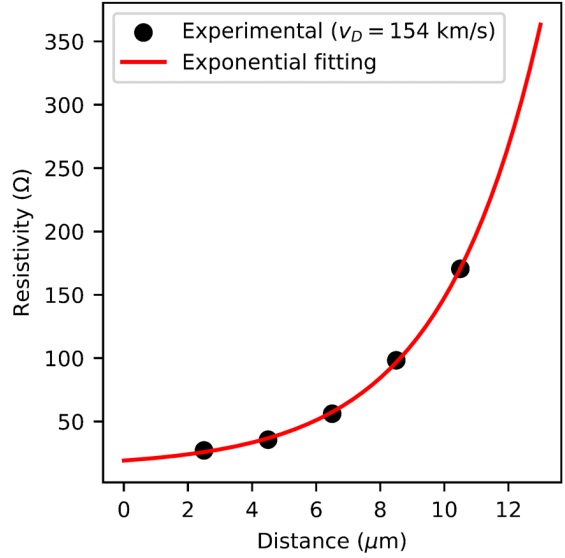

**Figure S2. Exponential fitting of the resistivity vs. distance curves.** The fitted parameters for the curve are: **$v_D$=154 km/s**, a=5.49±1.13 Ω, b=0.32±0.02 μm$^{-1}$, m=13.68±3.33 Ω.

### 1.3 Carrier density dependence of the differential resistivity for different pairs of contacts.

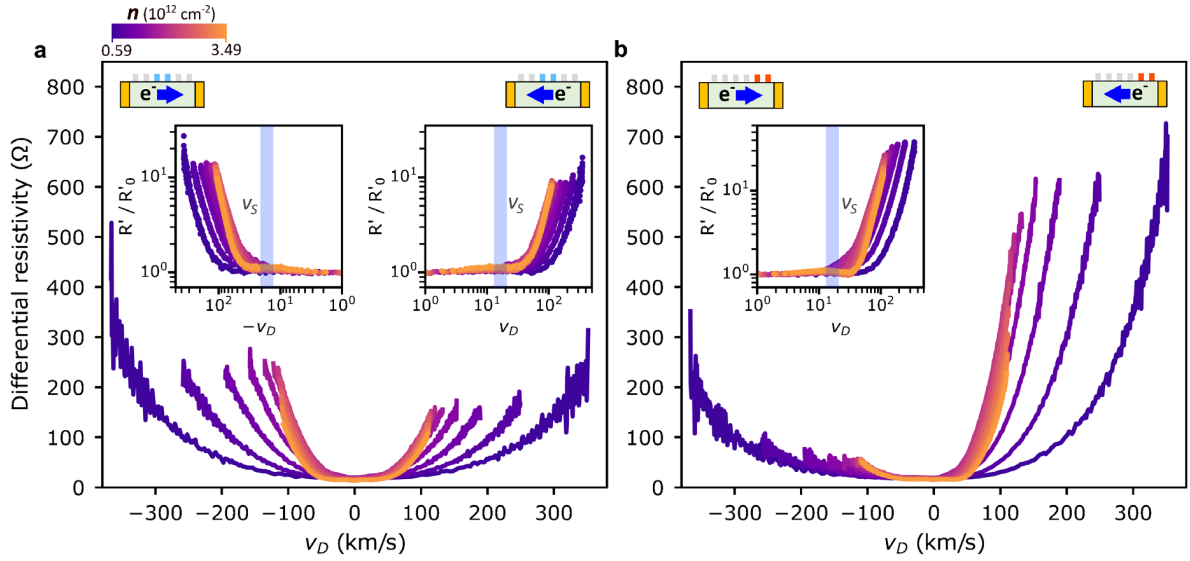

**Figure S3. Resistivity vs. drift velocity at different carrier densities for contacts 3-4 (a) and 5-6 (b).** Panel a and b show the differential resistivity vs. $v_D$ at different carrier concentrations for contacts 3-4 and 5-6 respectively of device A. For both pairs of contacts, a sharp transition between ohmic and non-ohmic behavior can be observed in the logarithmic insets (differential resistivity normalized to the value at vD=0). The shadowed regions indicate drift velocities between 13 and 21 km/s, corresponding to the speed of sound for TA and LA phonons respectively. The device schemes indicate the carrier flow direction, the type of carriers and the contacts being measured in each case (colored contacts).

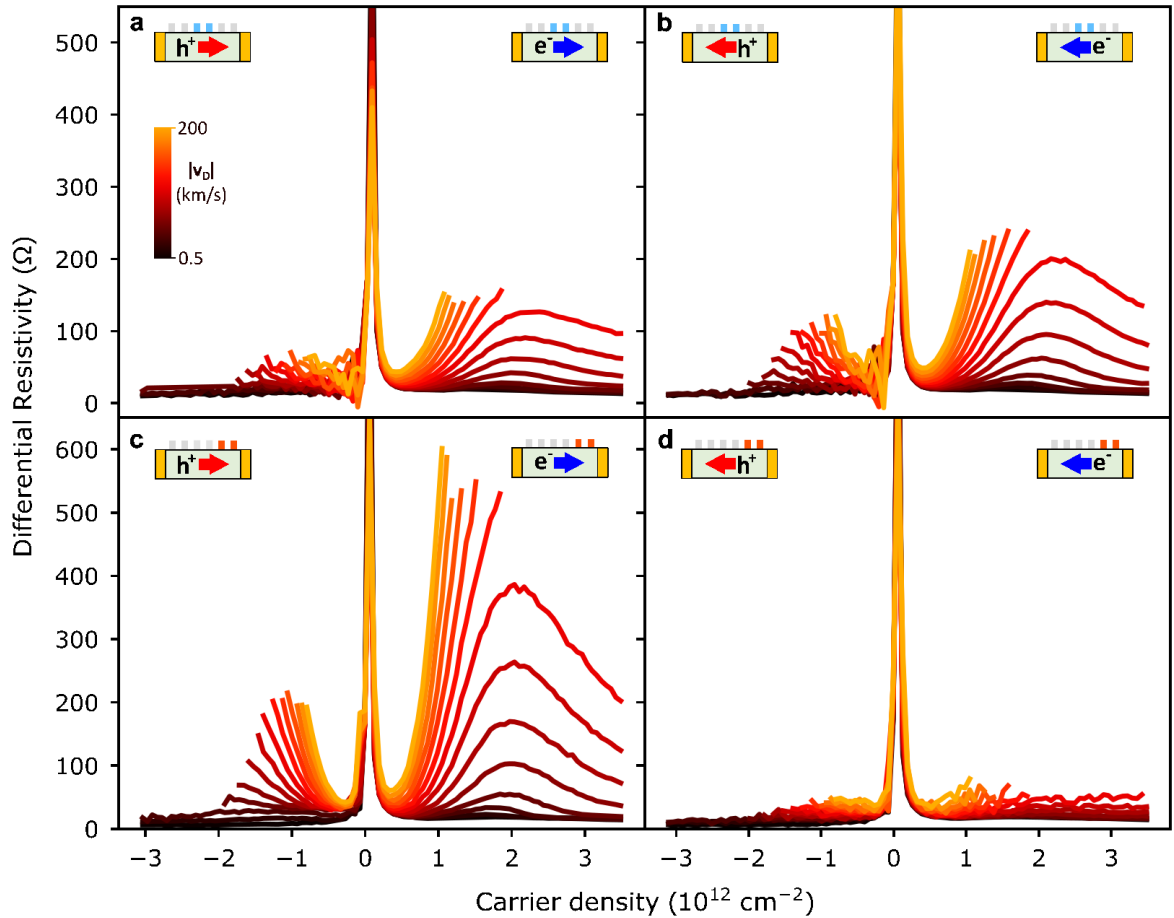

**Figure S4. Resistivity vs. Carrier density for contacts 3-4 and 5-6.** The top panel (a, b) shows the differential resistivity vs. carrier concentration for contacts 3-4 in both directions of the carrier flow. Since these contacts are located at the center of the device, phonon amplification is observed for both directions, however, the resistivity growth is not as large as in contacts 1-2 or 5-6. The bottom panel shows the case for contacts 5-6. As can be observed, just like in the case of contacts 1-2, there is large growth of the differential resistivity when the carriers flow downstream (c). On the other hand, the resistivity growth is barely noticeable if the direction of the current is reversed (d). The device cartoons indicate the carrier flow direction, the type of carriers and the contacts being measured in each case (colored contacts).

### 1.4 Temperature dependence for contacts 5-6.

Figure S5 shows the temperature dependence for contacts 5-6 at $n=1.4\times10^{12}$ cm$^{-2}$. The differential resistivity vs. $v_D$ at different temperatures (a) and resistivity vs. temperature curves (b) are similar to those of contacts 1-2, except that in this case the resistivity growth occurs for right-moving carriers. In Figure S5a, the asymmetry of the differential resistivity vs. $v_D$ curve at 280 K is still evident, just as observed for contacts 1-2 (Fig. 3a).

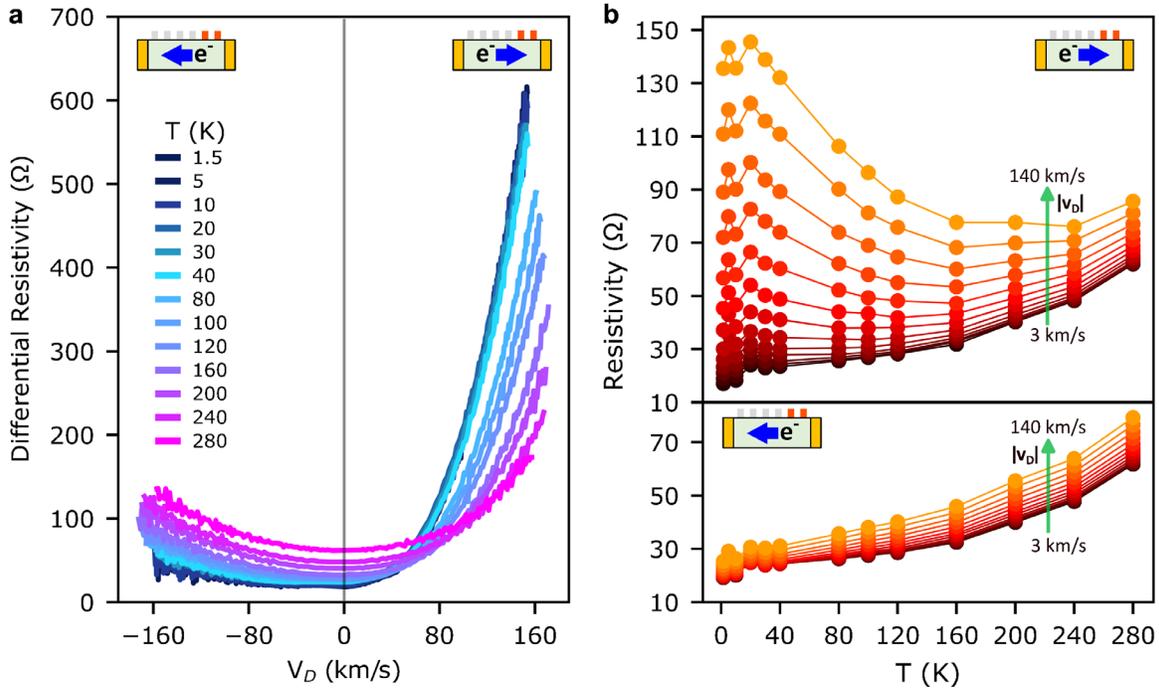

**Figure-S5. Temperature dependence at n=1.4×10$^{12}$ cm$^{-2}$ for contacts 5-6.** a) Differential resistivity vs. $v_D$ at different temperatures. b) Resistivity vs. T at different $v_D$ values.

## 2. Phonon amplification in Device B, second non-aligned device.

Device B is the second unaligned device presented in this work. Like device A, it also exhibits clear signs of phonon amplification (Figures S6, S7 and S8). First, we determine that the graphene is unaligned with the hBN from resistivity versus carrier density measurements which show a typical graphene Dirac peak (Figure S6b). When biased, we observe the same qualitative features measured in Device A presented in the main text figures:

1. The V-I curves are highly asymmetric due to the local resistivity growing in the carrier flow direction when the drift velocity exceeds the sound velocity (Figure S6).
2. The onset of non-ohmic behavior occurs with a sharp threshold only above the sound velocity (Figure S7b).
3. The resistance growth occurs away from the Dirac point and increases with carrier density (Figure S7c). Note, unlike in Device A, we do not observe the resistance growth decreasing again at higher carrier densities.
4. The asymmetric resistance growth is enhanced at low temperatures. The excess resistivity induced by biasing at low temperatures is larger than that induced by heating to room temperature (Figure S8).

A difference is that device B is clearly more disordered than device A, which we observe as a spatially varying resistance across the device channel at low biases that does not depend on the direction of the current. As a result, the mobility measured in contacts 1-2 has a value of 20.8 m$^2$/V*s whereas for contacts 4-5 it decreases to 1.25 m$^2$/V*s. This we ascribe to increased disorder closer to the drain contact. The effect that such disorder has on the resistance growth can also be understood within the phonon amplification model. In Figure S6d, it is shown that the resistivity growth with distance at different $|v_D|$ values, is greater when the carriers move from source to drain (left panels). This suggests that phonon amplification is inhibited when the carriers move in the opposite direction, which supports the claim that device B is more disordered close to the drain contact.

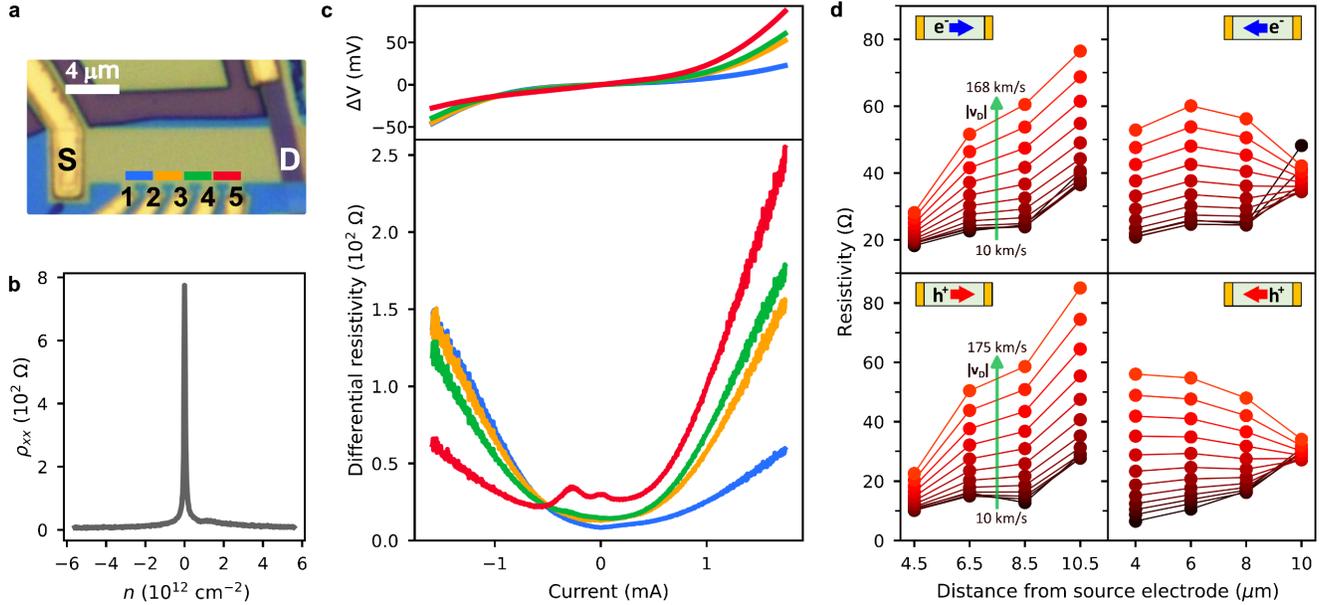

**Figure-S6. Device B, V-I curves and resistivity distance dependence. a)** Optical image of device B with dimensions L=14.5 µm, w=4 µm and center to center distance between voltage tabs of 2 µm. The color code in this image is used to identify the line traces corresponding to each pair of contacts in panel c, S and D indicate the source and drain electrodes respectively. **b)** Dirac peak for device B, the absence of satellite peaks indicates that the graphene is not aligned with any of the hBN layers. **c)** Raw V-I curves (top) and differential resistivity vs. source-drain current (bottom) for all the pairs of contacts in device B at n=-1.4×10$^{12}$ cm$^{-2}$ (hole doped, Vg=-21 V). **d)** Length dependence of the differential resistivity at n=1.4×10$^{12}$ cm$^{-2}$ (top panel) and n=-1.4×10$^{12}$ cm$^{-2}$ (bottom panel). The device cartoons in each plot show the flow direction and type of carriers for each case. The green arrows indicate the minimum and maximum drift velocities for top and bottom panels. All measurements are performed at T= 2.8 K.

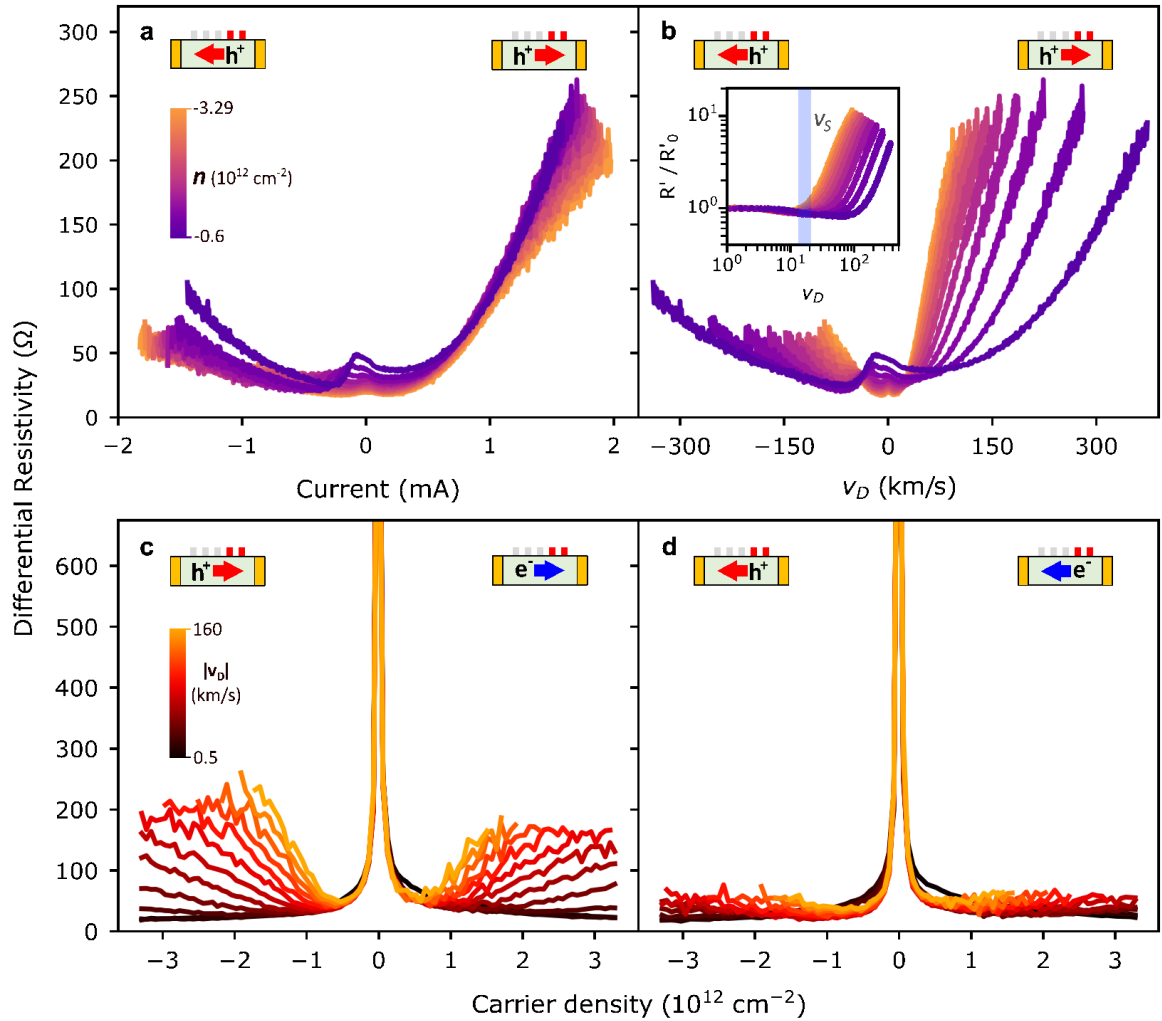

**Figure S7. Carrier density dependence for device B, contacts 4-5.** Top: Differential resistivity as a function of **a)** current and **b)** drift velocity at different carrier densities, inset: normalized differential resistivity. Bottom: Differential resistivity vs. carrier density for **c)** downstream and **d)** upstream carrier flow. The device cartoons indicate the carrier flow direction, type of carriers and contacts being measured (colored contacts).

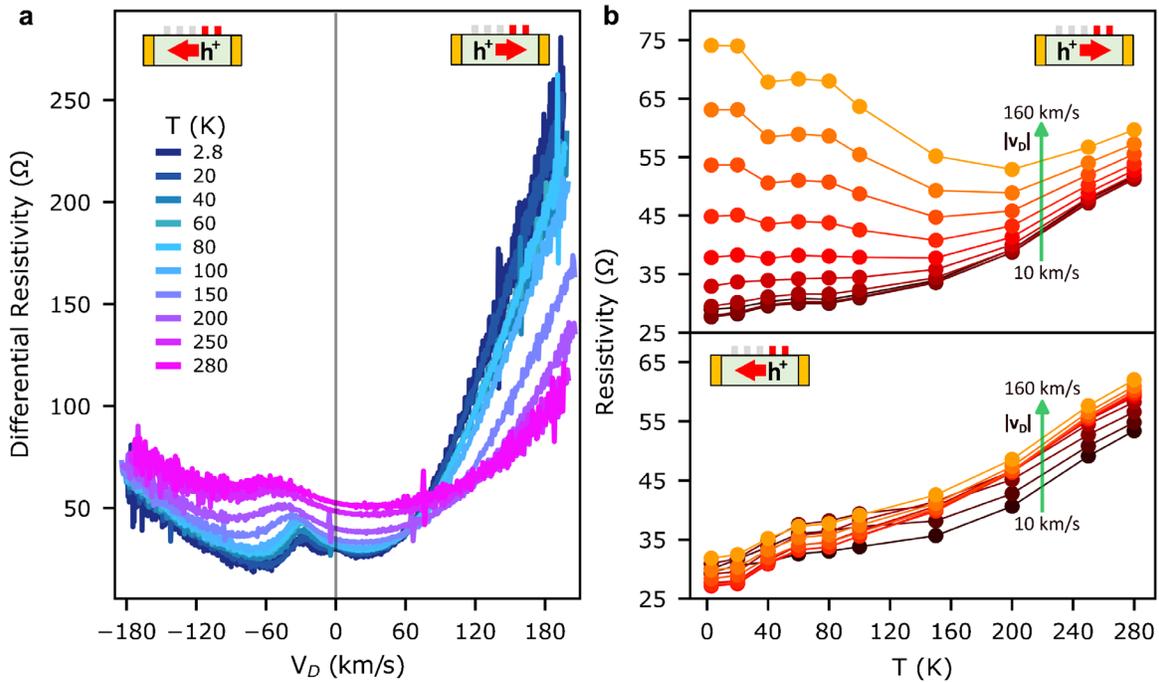

**Figure S8. Temperature dependence of device B at n=-1.4×10$^{12}$ cm$^{-2}$ for contacts 4-5.** a) Differential resistivity vs. v$_D$ curves at different temperatures. b) Resistivity vs. T at different v$_D$ values.

### 3. Device C: Graphene/hBN aligned device.

Figure S9 shows the measurements taken for device C, which is fabricated identically to devices A & B, but has the graphene aligned to at least one of the hBN layers. Evidence for alignment comes from the resistivity vs. carrier density trace, which shows clear satellite resistance peaks and an overall higher resistivity of the device (Figure S9b). This indicates that at least one of the hBN layers is aligned at some low angle to the graphene layer, producing a 14 nm length scale moiré. This aligned device C shows extremely different transport behavior than the non-aligned devices A & B. The differential resistivity vs. current curves are symmetric, and do not show the characteristic asymmetry seen in non-aligned devices that is the hallmark of phonon amplification (Figure S9c). Likewise, the spatial dependence of the differential resistivity does not depend on the direction of the current (Figure S9d), and likely arises from inherent sample inhomogeneity, such as twist angle variations. In the case of this aligned device, the nonlinearity of the V-I curve likely arises from Joule heating and Umklapp scattering[1]. In summary, there is no evidence of phonon amplification occurring in the aligned device C.

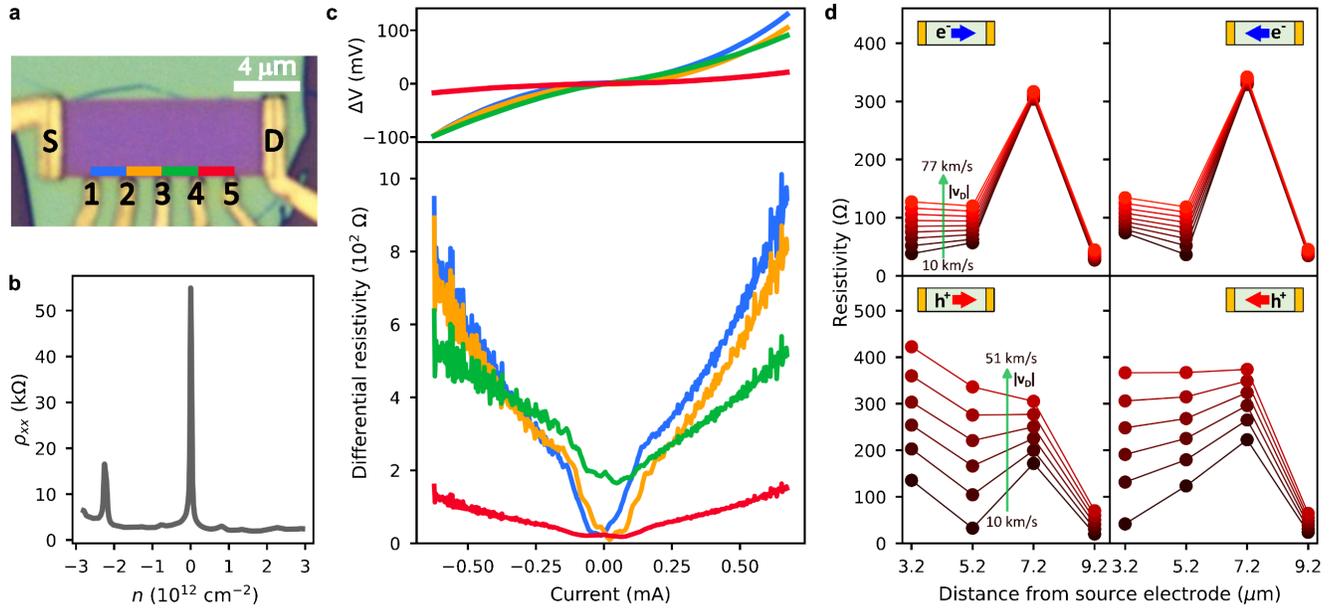

**Figure-S9. Device C. a)** Optical image of device C, the color code in this image is used to identify the line traces corresponding to each pair of contacts in panel c, S and D indicate the source and drain electrodes respectively. **b)** Dirac peak for device C, the satellite peaks indicate that the graphene is aligned with any of the hBN layers. **c)** Raw V-I curves (top) and differential resistivity vs. source-drain current (bottom) for all the pairs of contacts in device C at n=-1.4×10$^{12}$ cm$^{-2}$ (hole doped, Vg=-25 V). **d)** Length dependence of the differential resistivity at n=1.4x10$^{12}$ cm$^{-2}$ (top panel) and n=-1.4×10$^{12}$ cm$^{-2}$ (bottom panel). The device cartoons in each plot show the flow direction and type of carriers for each case. All measurements are performed at T = 1.5 K.

## 4. Device D: Disordered device.

Device D is fabricated identically to A and B but exhibits 10x lower mobility (at n=1.4×10$^{12}$ cm$^{-2}$ and T=2.8 K), which varies across the device from 1.8 m$^2$/V*s to 2.78 m$^2$/V*s for contacts 1-2 and 3-4 respectively. The device shows substantial variation in the local resistance, as well as an optically non-uniform appearance which suggests the presence of impurities from the fabrication process. It does not exhibit any signs of graphene-hBN alignment. The V-I curves show only weak nonlinearities that are symmetric with current direction. The lack of resistance growth in this disordered sample supports the phonon amplification model, where sample disorder would be expected to scatter phonons and inhibit the amplification process.

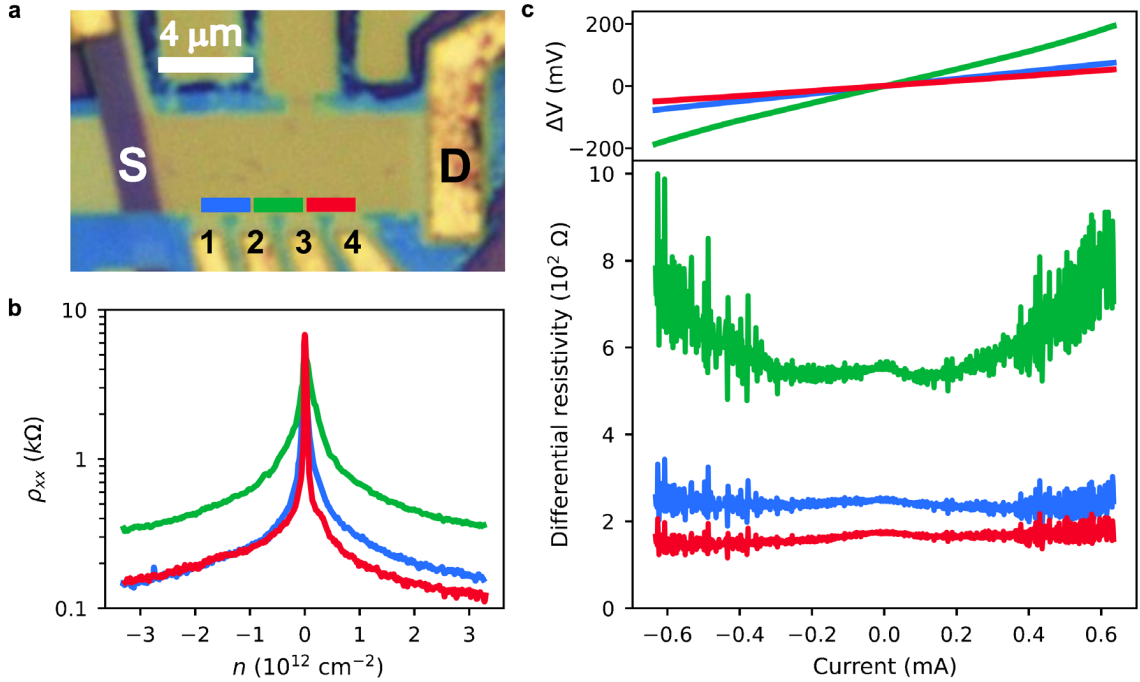

**Figure S10. Device D: a)** Optical image of device D, the color code in this image is used to identify the line traces corresponding to each pair of contacts in panels b and c. **b)** Logarithmic scale Dirac peaks for the different pairs of contacts in device D. **c)** Raw V-I curves (top) and differential resistivity vs. source-drain current (bottom) for all the pairs of contacts in device D at n=1.4×10$^{12}$ cm$^{-2}$ (electron doped, Vg=21 V). All measurements are performed at T = 2.8 K.

## 5. Theory calculations.

Our calculation works in three steps. First, we assume a drifting Fermi-Dirac distribution with drift velocity $v_d$ for the electrons in order to calculate the phonon amplification rate $\Gamma_{\mathbf{q}}^{\mathrm{amp}}$ due to electron-phonon coupling. Second, we use $\Gamma_{\mathbf{q}}^{\mathrm{amp}}$ to calculate the position-dependent out-of-equilibrium phonon distribution in the sample. Finally, we find the (position-dependent) electric field needed to sustain the assumed $v_d$ based on the electronic Boltzmann equation in which the phonon distribution enters through the electron-phonon scattering integral. Our calculation is intended to reproduce the main physical features of the system with a minimum amount of details, and as such has limitations. The main one is our assumption of a drifting Fermi-Dirac distribution for the electrons, with a nominal temperature which is $v_d$-independent. In the actual system, we would expect that Joule heating would lead to increased electronic temperatures at high $v_d$, which could be considered in a refined calculation which treats electrons and phonons in a self-consistent way. Such a calculation is left for future work.

## 5.1 Phonon amplification rate.

Phonons with a given wavevector **q** are amplified by the electron-phonon interaction with a characteristic amplification rate[2]:

$$\Gamma_{\mathbf{q}}^{\text{amp}} = \gamma_{\mathbf{q}}^{\text{em}} - \gamma_{\mathbf{q}}^{\text{abs}} - \tau_{\mathbf{q}}^{-1}$$

$$= \left[\frac{2\pi}{\hbar} g_s g_v \sum_{\mathbf{k}} \sum_{\mathbf{k'}} |C_{\mathbf{k},\mathbf{k'},\mathbf{q}}|^2 \cdot [f_{\mathbf{k}}(1-f_{\mathbf{k'}}) - (1-f_{\mathbf{k}})f_{\mathbf{k'}}] \cdot \delta_{\mathbf{k}-\mathbf{k'},\mathbf{q}} \cdot \delta(E_{\mathbf{k}} - E_{\mathbf{k'}} - \hbar\omega_{\mathbf{q}})\right] - \tau_{\mathbf{q}}^{-1},$$

where:

$$|C_{\mathbf{k},\mathbf{k'},\mathbf{q}}|^2 = \frac{D^2 \hbar |\mathbf{q}|}{2\rho A v_s} \cos^2\left(\frac{\theta_{\mathbf{k},\mathbf{k'}}}{2}\right),$$
$$E_{\mathbf{k}} = \hbar v_F |\mathbf{k}|,$$
$$\omega_{\mathbf{q}} = v_s |\mathbf{q}|,$$

and $f_{\mathbf{k}}$ is the drifting Fermi-Dirac distribution given by:

$$f_{\mathbf{k}} = \frac{1}{1 + \exp[\beta(E_{\mathbf{k}} - \hbar v_d k_x - \mu)]},$$

where we have chosen $v_d \parallel \hat{x}$. We choose to use the deformation potential $D = 19$ eV [3]. $A$ is the sample area, and $\rho = 7.63 \times 10^{-8}$ g/cm$^2$ is the mass area density of graphene. The spin degeneracy $g_s$ and valley degeneracy $g_v$ are both 2. The chemical potential $\mu$ is related to the carrier density $n$ by $\mu = \hbar v_F \sqrt{\pi n}$, and the Fermi velocity $v_F = 1000$ km/s. To simplify our computation, we consider only the longitudinal acoustic phonons, for which $v_s = 21$ km/s $= 0.021 v_F$. Phonon decay from other sources, such as the anharmonic interaction or point-defect scattering, are incorporated into $\tau_{\mathbf{q}}^{-1}$. The other two terms tend to be larger by at least an order of magnitude[2], so for simplicity we ignore the **q**-dependence and choose $\tau_{\mathbf{q}}^{-1} = 100$ MHz. Plots of $\Gamma_{\mathbf{q}}^{\text{amp}}$ for various drift velocity appear in Figure S11a, and line cuts along $q_y = 0$ appear in Figure S11b. $\Gamma_{\mathbf{q}}^{\text{amp}}$ is positive for $v_d > v_s$ in a cone along the direction of $v_d$. Above $v_s$, the strength of the effect increases linearly with $v_d$, as shown in Figures S11 and S12b. Additionally, the strength of the effect scales linearly with the carrier density $n$, since the summation in the computation for $\Gamma_{\mathbf{q}}^{\text{amp}}$ gives an overall factor of $k_F^2$. This is shown in Figure S11c.

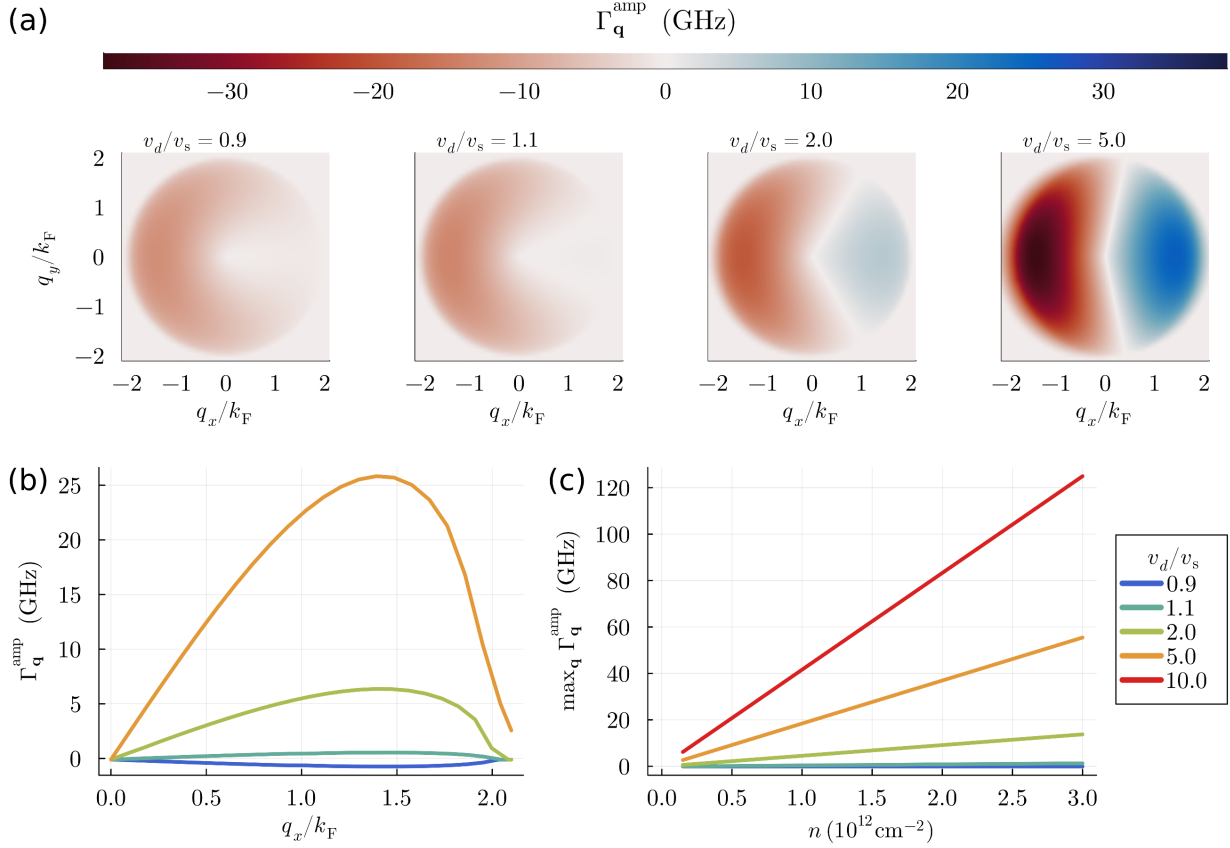

**Figure S11. Phonon amplification rates. a)** The amplification rate $\Gamma_{\mathbf{q}}^{\text{amp}}$ is plotted here for different values of $v_d$, with T = 2 K, $n = 1.4 \times 10^{12}$ cm$^{-2}$. The effect from the electron-phonon interaction is nonzero out to $|\mathbf{q}| = 2k_F$, with this cutoff being smoothed out at nonzero temperature. Positive amplification rates appear in a cone in the direction of $v_d$ once it exceeds $v_s$. The width of this cone and the magnitude of the effect increase as $v_d$ increases. **b)** $\Gamma_{\mathbf{q}}^{\text{amp}}$ line cuts along $q_y = 0$ for various drift velocities, T = 2 K, $n = 1.4 \times 10^{12}$ cm$^{-2}$. For $v_d < v_s$, the amplification rate from electron-phonon coupling is always negative, and thus $\Gamma_{\mathbf{q}}^{\text{amp}}$ never exceeds $-\tau_{\mathbf{q}}^{-1} = -100$ MHz. For $v_d > v_s$, electron-phonon coupling gives a positive amplification rate along the positive-$q_x$ axis out to $q_x = 2k_F$. The overall scale of $\Gamma_{\mathbf{q}}^{\text{amp}}$ increases as $v_d$ increases; for $v_d > v_s$, the maximum value of $\Gamma_{\mathbf{q}}^{\text{amp}}$ increases linearly with $v_d$ (also seen in Figure S12b). **c)** The maximum value of $\Gamma_{\mathbf{q}}^{\text{amp}}$ (maximized over $\mathbf{q}$) as a function of carrier density $n$ for various drift velocities, T = 2 K. At a given value of $n$, the maximum value of $\Gamma_{\mathbf{q}}^{\text{amp}}$ is larger for larger $v_d$. For a given $v_d$, the maximum value of $\Gamma_{\mathbf{q}}^{\text{amp}}$ scales linearly with $n$, since there is an overall factor of $k_F^2$ in the summation for $\gamma_{\mathbf{q}}^{\text{em}}$ and $\gamma_{\mathbf{q}}^{\text{abs}}$.

### 5.2 Phonon distribution.

Using the amplification rate $\Gamma_{\mathbf{q}}^{\text{amp}}$, we solve the Boltzmann transport equation for the phonons to obtain the $x$-dependent phonon population for each mode $\mathbf{q}$; for right-moving phonons this is:

$$n_{\mathbf{q}}(x) = n_{\mathbf{q},0} e^{\Gamma_{\mathbf{q}}^{\text{amp}} x/v_{\mathbf{q}}} + \frac{1}{\Gamma_{\mathbf{q}}^{\text{amp}} \tau_{\mathbf{q}}}\left(n_{\mathbf{q},0} + \tau_{\mathbf{q}} \gamma_{\mathbf{q}}^{\text{em}}\right)\left(e^{\Gamma_{\mathbf{q}}^{\text{amp}} x/v_{\mathbf{q}}} - 1\right),$$

and for left-moving phonons,

$$n_\mathbf{q}(x) = n_{\mathbf{q},0} e^{-\Gamma_\mathbf{q}^{amp}(x-L)/v_\mathbf{q}} + \frac{1}{\Gamma_\mathbf{q}^{amp}\tau_\mathbf{q}}(n_{\mathbf{q},0} + \tau_\mathbf{q}\gamma_\mathbf{q}^{em})\left(e^{-\Gamma_\mathbf{q}^{amp}(x-L)/v_\mathbf{q}} - 1\right),$$

where $v_\mathbf{q} = v_s \frac{q_x}{|\mathbf{q}|}$, $L$ is the length of the sample, and $n_{\mathbf{q},0}$ is the equilibrium Bose-Einstein distribution.

### 5.3 Resistivity.

Next, we compute the electric field $E$; to do this, we obtain a current-balance equation by multiplying both sides of the Boltzmann transport equation[4] for the electrons by $v_x$ and summing over $\mathbf{k}$,

$$\frac{-eE}{\hbar}\sum_\mathbf{k} v_x \frac{\partial f_\mathbf{k}}{\partial k_x} = g_s g_v \sum_\mathbf{k}\sum_{\mathbf{k}'} v_x [F(\mathbf{k},\mathbf{k}') - F(\mathbf{k}',\mathbf{k})],$$

where the scattering rate $F(\mathbf{k},\mathbf{k}')$ is given by[3]

$$F(\mathbf{k},\mathbf{k}') = f_\mathbf{k}(1 - f_{\mathbf{k}'})W(\mathbf{k},\mathbf{k}'),$$

$$W(\mathbf{k},\mathbf{k}') = \frac{2\pi}{\hbar}\sum_\mathbf{q} \frac{D^2\hbar|\mathbf{q}|}{2\rho A v_s}\cos^2\left(\frac{\theta_{\mathbf{k},\mathbf{k}'}}{2}\right)\delta_{\mathbf{k},\mathbf{k}'+\mathbf{q}}[(n_\mathbf{q}+1)\delta(E_\mathbf{k} - E_{\mathbf{k}'} - \hbar\omega_\mathbf{q}) + n_{-\mathbf{q}}\delta(E_\mathbf{k} - E_{\mathbf{k}'} + \hbar\omega_\mathbf{q})].$$

The resistivity is then computed as $E/j$, where $j = nev_d$. A plot of the resistivity versus $x$ for various $v_d$ appears in Figure S12a. We find that the resistivity is an exponential function of $x$, with the growth rate approximately corresponding to the maximum phonon amplification rate. For each $v_d$, we fit the logarithm of the resistance values versus $x$ using a linear function to determine the growth rate. For $v_d > v_s$, this growth rate closely matches $\max_\mathbf{q} \Gamma_\mathbf{q}^{amp}/v_s$, as shown in Figure S12b.

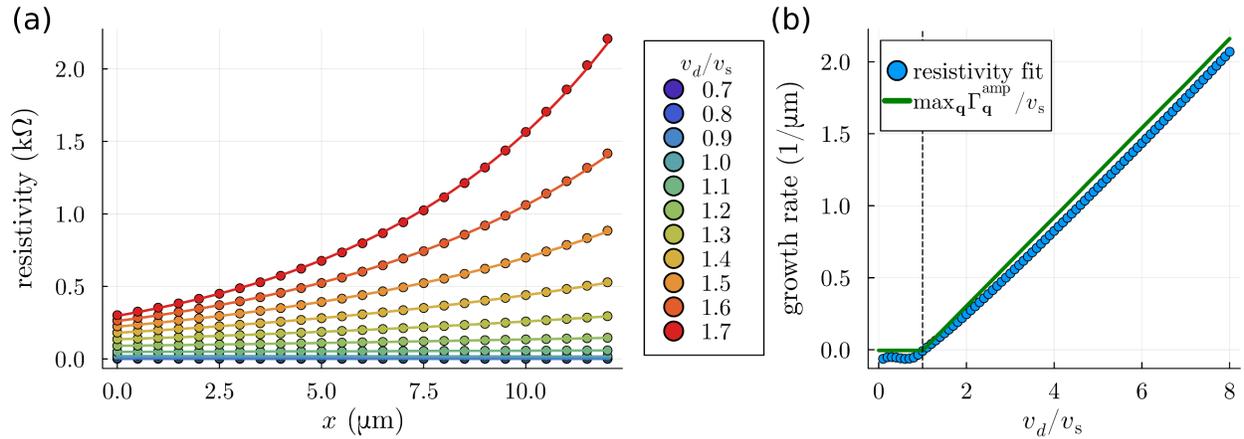

**Figure S12. a)** Resistivity versus $x$ for various $v_d$, $n = 1.4 \times 10^{12}$ cm$^{-2}$, $T = 2$ K, with exponential fits. **b)** Resistivity growth rate (blue dots) as a function of $v_d$. The green line is the maximum $\Gamma_q^{amp}$ divided by $v_s$ for each $v_d$, which agrees well with the growth rates determined from fitting the computed resistivities. The dashed black line shows $v_d = v_s$.